\documentclass[12pt,a4paper]{article}
\usepackage{amsmath}
\usepackage{amsfonts}
\usepackage{setspace}
\usepackage{array}
\usepackage{hyperref}
\usepackage{cite}
\usepackage{amssymb}
\usepackage{graphicx}
\usepackage{float}

\providecommand{\keywords}[1]{\textbf{\textit{Keywords--}} #1}
\title{Active-Sterile neutrino masses and mixings in $A_4$ minimal extended seesaw mechanism}

\author{M. Kishan Singh$^1\footnote{\footnotesize kishan@manipuruniv.ac.in}$, S. Robertson$^1\footnote{\footnotesize robsoram@gmail.com}$, N. Nimai Singh$^{1,2}\footnote{\footnotesize nimai03@yahoo.com}$ \\ \small{$^1$Department of Physics, Manipur University, Imphal-795003, India.}\\
\small{$^2$Research Institute of Science and Technology, Imphal-795003, India.}}

\date{}

\begin{document}

\maketitle

\begin{abstract}
Assuming the existence of an eV or KeV scale sterile neutrino, we develop a 3+1 neutrino mass model using $A_4\times Z_4 \times Z_2$ symmetry group. Three Higgs $H, H^{\prime}$ and $H^{\prime\prime}$ are considered to give a desired neutrino mass matrix which generates non-zero $\theta_{13}$. The model can give neutrino mixing parameters that are well within the experimental 3$\sigma$ bounds. We also calculate the $4\times 4$ active-sterile neutrino mixing matrix, and it is found to be consistent with experimental bounds.
\end{abstract}


\newpage
\section{Introduction}  

The theory of neutrino masses and mixings has been a very highly exciting field of research in recent years. Observations from SNO, T2K, etc., that neutrinos have masses and oscillate among different flavors, have triggered a new approach in the study of neutrinos beyond the standard model (SM) framework. The three generations of leptonic mass squared differences and mixing angles have reached the precision measurement status. However, other parameters such as mass hierarchy,absolute neutrino mass, Dirac CP-violating phase are still unknown. The current global fit data for three neutrino oscillations\cite{2020data} are shown in Table \ref{Table1}.

Apart from these, one interesting anomaly came from the experimental data of LSND \cite{aguilar2001evidence}, which observed  excess of electron anti-neutrino $(\overline{\nu}_e)$ in a muon anti-neutrino $(\overline{\nu}_\mu)$ beam produced at the Los Alamos laboratory. Another experiment called, MiniBooNE\cite{aguilar2018significant} supplemented LSND results and  observed an oscillation $\overline{\nu}_\mu $ to $\overline{\nu}_e$ compatible with the LSND data. These results can be interpreted by including more mass eigenstates of neutrino in the three neutrino theory\cite{Dentler}. One of the simplest ways is to add a fourth neutrino state, generally called the sterile neutrino, to the three active neutrinos\cite{Zhang2011}. The word sterile refers to the fact that such a neutrino can not have weak interactions in the SM from the requirement of being in agreement with the precision measurement of Z boson decay width at the LEP experiment. However, they can mix with the active neutrinos. It has been shown\cite{giunti2020new} in the 3+1 framework that the new analysis of the MiniBooNE data allows smaller values of active-sterile neutrino mixing for the standard analysis. Super-Kamiokande (SK) has provided upper bounds on sterile neutrino parameter $|U_{\tau 4}|< 0.18$  at 90$\%$ CL \cite{SK}. IceCube Collaborations\cite{Icecube2017search} has also analyzed light sterile neutrinos from three years of atmospheric neutrino data from the DeepCore detector and provided limits on sterile neutrino mixing at  $|U_{\mu 4}|^2 <0.11$ and $|U_{\tau 4}|^2<0.15$ (90$\%$ C.L.) for the sterile neutrino mass splitting $\Delta m_{41}^2 =1.0$ eV$^2$. New data from the reactor and other short and long-baseline neutrino experiments such as MINOS \cite{minos}, Daya Bay \cite{dayabay} etc., provide new bounds on active-sterile mixing and $\Delta m_{41}^2$. Recently, it has been reported from the MicroBooNE that there is no hint of an eV-scale sterile neutrino. But, GeV to KeV scale sterile neutrinos are still well-motivated theoretically and do not contradict any existing experiments. Besides, MicroBooNE does not probe the full parameter space of sterile neutrino models hinted at by MiniBooNE and other data, nor do they probe the $\nu_e$ interpretation of the MiniBooNE excess in a model-independent way\cite{2021microboone}.  Many other ongoing and future long-baseline experiments such as DUNE \cite{Dune}, T2HK \cite{t2hk}, T2HKK \cite{t2hkk} etc. may provide new insights on neutrino oscillation physics and explore active-sterile mixing.  The phenomenology and experimental constraints on (3+1) neutrinos have been reviewed in \cite{kisslinger2014sterile,DEV2019401, barry2011light,gariazzo2016light,zhang2013naturally,zhang2013majorana,
borah2017nonzero,Das1,Das2,Das3}.

\begin{table}
\centering
\renewcommand{\arraystretch}{1.5}
\begin{tabular}{|c|c|c|c|}
 \hline 
parameter &	best fit $\pm$ 1$\sigma$ &	2$\sigma$ range &	3$\sigma$ range \\
\hline
$|\Delta m^2_{21}|: [10^{-5} eV^2]$ &	$7.50^{+0.22}_{-0.20}$ &	7.11–7.93 &	6.94–8.14 \\
$|\Delta m^2_{31}|: [10^{-3} eV^2] (NO)$	& $2.55^{+0.02}_{-0.03}$ & 2.49–2.60 & 2.47–2.63 \\
$|\Delta m^2_{32}|: [10^{-3} eV^2] (IO)$ &	$2.45^{+0.02}_{-0.03}$ & 2.39–2.50 & 2.37–2.53\\
$\sin^2\theta_{12}/ 10^{-1}$	& $3.18\pm 0.16$&	2.86–3.52 & 2.71–3.69 \\
$\sin^2\theta_{23}/ 10^{-1} (NO)$ &	$5.74\pm0.14$ &	5.41–5.99 &	4.34–6.10 \\

$\sin^2\theta_{23}/ 10^{-1} (IO)$ &	$5.78^{+0.10}_{-0.17}$ &	5.41–5.98	&4.33–6.08 \\

$\sin^2\theta_{13}/ 10^{-2} (NO)$ &	$2.200^{+0.069}_{-0.062}$ & 2.069–2.337	&2.000–2.405 \\

$\sin^2\theta_{13}/ 10^{-2} (IO)$ &	$2.225^{+0.064}_{-0.070}$ &	2.086–2.356	&2.018–2.424 \\

$\delta_{\rm CP}/\pi (NO)$ &	$1.08^{+0.13}_{-0.12}$	& 0.84–1.42 &	0.71–1.99 \\

$\delta_{\rm CP}/\pi (IO)$ & $1.58^{+0.15}_{-0.16}$ &	1.26–1.85 &	1.11–1.96 \\
\hline
\end{tabular}  
\caption{ Updated global-fit data for three neutrino oscillation\cite{2020data}.}
\label{Table1}
\end{table} 

Many authors have used discrete flavor symmetries of order $n$ such as $A_n,\ S_n,\ C_n,\ Z_n$, etc., to develop models consistent with the current oscillation data. Several mechanisms have been used to study possible active-sterile mixing within the seesaw models. In Ref. \cite{Dasthesis}, the authors have extensively studied minimal extended seesaw mechanism (MES) and their possible effects on cosmological problems such as baryogenesis, neutrinoless double beta decay, dark matter, etc. In the present work, we will use the MES mechanism to explain active-sterile mixing. We use $A_4,~Z_4$ with three Higgs (one SM Higgs and two BSM Higgs) doublets with three $A_4$ triplet and three $A_4$ singlet flavon fields to explain possible active-sterile mixing in the normal heirarchy (NH). We take $Z_2$ symmetry group to remove unwanted interactions in the Lagrangian. We give different group charges to the fields resulting to new interactions and therefore, generate different neutrino mass matrix structures. We develop a $(4\times 4)$ active-sterile neutrino mixing matrix from the model consistent with current experimental data. Moreover, there are searches for massive neutrinos, such as kinematic measurements of $\beta-$decay and searches for neutrinoless double beta decay $(0\nu\beta\beta)$ events. The absolute neutrino mass scale is directly probed from the cut-off of the electron energy spectrum emitted from $\beta-$decay\cite{Hagstot}. The effective neutrino mass $m_{\beta}$ is given as 
\begin{equation}
m_{\beta} = \left(\sum_{i=1}^4|U_{ei}|^2m_{i}^2\right)^{1/2},
\label{mbeta}
\end{equation}  
where $U$ is the $4\times 4$ active-sterile neutrino mixing matrix.   The effective mass parameter $m_{\beta\beta}$ from $0\nu\beta\beta$ can be expressed as the sum of mass eigenstates and mixing matrix elements as 
\begin{equation}
m_{\beta\beta}=\vert\sum_{j=1}^4|U_{ej}|^2m_j\vert.
\end{equation} 
We solve these effective mass parameters from the model. The outline of this paper is as follows. We have a brief discussion about the MES mechanism in section 2, followed by the description of our model in section 3. In section 4, we will show the numerical analysis of the neutrino masses and mixing matrices and the results of our work. We will conclude with a summary and discussion in section 5.

\section{Minimal extended seesaw mechanism}
In this mechanism, along with the SM particles, we take three extra right-handed singlet neutrinos ($\nu_{R1}, \nu_{R2}, \nu_{R3}$) and one additional gauge singlet chiral field $S$. It is possible to naturally generate an eV scale or KeV scale sterile neutrino mass with minimal but non-zero mixing with active neutrinos. 

The general Lagrangian of neutrino mass terms is given by 
\begin{equation}
-\mathcal{L}=\overline{\nu_L}M_D\nu_R + \frac{1}{2}\overline{\nu_R^c}M_R\nu_R + \overline{S^c}M_S\nu_R + h.c.
\end{equation}
where $M_D$ and $M_R$ are the Dirac and Majorana mass matrices respectively. $M_S$ is a $(1\times 3)$ sterile neutrino mass matrix arising from the inclusion of only one extra singlet $S$. In the basis $(\nu_L,\nu_R^c,S^c)$, we get a full $(7\times 7)$ matrix given by\cite{Zhang2011} 
\begin{equation}
M_{\nu}^{7\times 7} = \left( \begin{matrix}
0 & M_D & 0 \\ 
M_D^T & M_R & M_S^T \\ 
0 & M_S & 0
\end{matrix} \right).
\end{equation}

In analogy to the canonical type-I seesaw, if we take the right-handed neutrino to be  much higher than the electroweak scale of $M_D$, they should be decoupled at low scales. Therefore, we can block-diagonalise the $(7\times 7)$ matrix by using the seesaw formula with the condition $M_R >> M_D $ and get a $(4\times 4)$ neutrino mass matrix in the basis $(\nu_L, S^c)$ as 

\begin{equation}
M_{\nu}^{4\times 4} = -\left(\begin{matrix}
M_DM_R^{-1}M_D^T & M_DM_R^{-1}M_S^T \\
 
M_S(M_R^{-1})^TM_D^T & M_SM_R^{-1}M_S^T
\end{matrix} \right).
\label{mv44}
\end{equation}

We have four light neutrino eigenstates corresponding to three active neutrinos and one sterile neutrino. As we can see in eq. (\ref{mv44}), $det(M_{\nu}^{4\times 4}) = 0.$ This means that at least one of the four light neutrinos is massless. We further proceed to diagonalize the above $4\times 4$ mass matrix with the seesaw condition that $M_D < M_S$, we obtain a leading order of the active neutrino mass matrix $m_{\nu}$ as well as the sterile mass $m_s$ given as follows.
\begin{equation}
m_{\nu} \simeq M_DM_R^{-1}M_S^T\left(M_S M_R^{-1}M_S^T\right)^{-1}M_S\left(M_R^{-1}\right)^T M_D^T-M_DM_R^{-1}M_D^T;
\label{mv}
\end{equation}
\begin{equation}
m_s \simeq - M_SM_R^{-1}M_S^T.
\label{ms}
\end{equation}

We can naturally produce a sterile neutrino having mass in the eV scale. For example, if we take $M_D \sim 10^2$ GeV, $M_R \sim 5 \times 10^{14}$ GeV and $M_S \sim 5 \times 10^2$ GeV , we get approximately $m_\nu \sim 0.02$ eV and $m_s \sim 0.5$ eV. Further, as pointed out in Ref. \cite{dasgupta}, a slightly higher keV scale sterile neutrino can also be generated if we increase the $M_S$ mass scale upto TeV scale, which is also possible in MES mechanism.

The $3\times 3$ active neutrino mass matrix $m_{\nu}$ can be diagonalized by a unitary $3\times 3$ complex matrix $U_{PMNS}$ as \cite{valle2006neutrino}
\begin{equation}
m_{\nu}=U_{PMNS}.diag(m_1,m_2,m_3).U_{PMNS}^T.
\label{u33}
\end{equation}
In MES scheme, for NH: $(m_1<<m_2<m_3<<m_4),$ the light neutrino masses including the eV or KeV mass scale sterile neutrino are given in terms of mass-squared differences as 
\begin{align*}
m_1=0\ ;\ \ \  m_2 =\sqrt{\Delta m_{21}^2}\ ;\ \ \ 
m_3=\sqrt{\Delta m_{21}^2 +\Delta m_{31}^2}\ ;\ \ \
m_4=\sqrt{\Delta m_{41}^2},
\end{align*}
where $\Delta m_{ij}^2 = |m_j^2-m_i^2|.$

$U_{PMNS}$ can be parameterized using three mixing angles $\theta_{12},\theta_{13}, \theta_{23}$ and one CP violating phase $\delta_{13}$ for Dirac neutrinos and two Majorana phases $\alpha$ and $\beta$ for Majorana neutrinos. In PDG convention, the general form of $U_{PMNS}$ is  
\begin{equation}
U_{PMNS}=\left(
\begin{matrix}
 1 & 0 & 0 \\
 0 & c_{23} & s_{23} \\
 0 & -s_{23} & c_{23} \\
\end{matrix}
\right)\left(
\begin{matrix}
 c_{13} & 0 & e^{-i\delta_{13}}s_{13} \\
 0 & 1 & 0 \\
 -e^{-i\delta_{13}}s_{13} & 0 & c_{13} \\
\end{matrix}
\right)\left(
\begin{matrix}
 c_{12} & s_{12} & 0 \\
 -s_{12} & c_{12} & 0 \\
 0 & 0 & 1 \\
\end{matrix}
\right)P,
\label{upmns}
\end{equation} 
where  $c_{ij}=\cos{\theta_{ij}},\ s_{ij}=\sin{\theta_{ij}}$ and $P = diag(1, e^{i\alpha},e^{i(\beta+\delta_{13})})$ is the Majorana phase matrix.

In the 3+1 mixing framework, the leptonic mixing matrix $U_{PMNS}$ is not strictly unitary due to contributions from the sterile sector. But, because of minimal active-sterile mixing, we can assume that $U_{PMNS}$ is unitary at the $\mathcal{O}(10^{-2})$ level \cite{Xing2020}.
The full $4\times 4$ neutrino mixing matrix takes the form \cite{v441982}
\begin{equation}
V \simeq \left(\begin{matrix}
(1-\frac{1}{2}RR^{\dagger})U_{PMNS} & R \\ 
-R^{\dagger}U_{PMNS} & 1-\frac{1}{2}R^{\dagger}R
\end{matrix} \right),
\label{V44}
\end{equation}
where R is a $3\times 1$ matrix which determines the strength of active-sterile mixing and 
\begin{equation}
R = M_DM_R^{-1}M_S^T(M_SM_R^{-1}M_S^T)^{-1}.
\end{equation}
Taking the same mass scales as shown above, we estimate $R \sim 0.2$ for $m_s$ in eV scale and $R \sim 0.1$ for $m_s$ in KeV scale, both of which are in good agreement with experimental data of active-sterile neutrino mixing.

The $4\times 4$ neutrino mixing matrix can also be parameterized by six mixing angles $(\theta_{12},\theta_{13}, \theta_{23},\theta_{14},\theta_{24}, \theta_{34})$, three Dirac phases $(\delta_{13},\delta_{14},\delta_{24})$ and three Majorana phases $(\alpha,\beta,\gamma)$\cite{gariazzo2016light}.
\begin{equation}
U^{4\times 4} = \left(\begin{matrix}
c_{12}c_{13}c_{14} & c_{13}c_{14}s_{12}e^{i\frac{\alpha}{2}} & c_{14}s_{13} e^{i\frac{\beta}{2}} & s_{14}e^{-i\frac{\gamma}{2}} \\ 
U_{\mu 1} & U_{\mu 2} & U_{\mu 3} & c_{14}s_{24}e^{-i\left(\frac{\gamma}{2}-\delta_{14}+\delta_{24}\right)} \\ 
U_{\tau 1} & U_{\tau 2} & U_{\tau 3} & c_{14}c_{24}s_{34}e^{-i\left(\frac{\gamma}{2}-\delta_{14}\right)} \\ 
U_{s1} & U_{s2} & U_{s3} & c_{14}c_{24}c_{34}e^{-i\left(\frac{\gamma}{2}-\delta_{14}\right)}
\end{matrix} \right).
\label{U44}
\end{equation}

Comparing eq. $(\ref{V44})$ and eq. $(\ref{U44})$, we obtain the relation between neutrino mixing angles and the elements of mixing matrix as 
\begin{align} \label{ang14}
\sin^2\theta_{14}\ \ & =\  |V_{e4}|^2 \\ 
\sin^2\theta_{24}\ \  &=\ \frac{|V_{e4}|^2}{1-|V_{e4}|^2} \\
Sin^2\theta_{34}\ \ &=\ \frac{|V_{\tau 4}|^2}{1-|V_{e4}|^2-|V_{\mu 4}|^2} \\
\sin^2\theta_{12}\ \ &=\ \frac{|V_{e2}|^2}{1-|V_{e4}|^2-|V_{e3}|^2} \\
\sin^2\theta_{13}\ \ &=\ \frac{|V_{e3}|^2}{1-|V_{e4}|^2} \\
\sin^2\theta_{23}\ \ &=\ \frac{|V_{e3}|^2(1-|V_{e4}|^2)-|V_{e4}|^2|V_{\mu 4}|^2}{1-|V_{e4}|^2-|V_{\mu 4}|^2} \\
 & +\ \frac{|V_{e1}V_{\mu 1}+V_{e2}V_{\mu 2}|^2(1-|V_{e4}|^2}{(1-|V_{e4}|^2-|V_{e3}|^2)(1-|V_{e4}|^2-|V_{\mu 4}|^2)}  \\ \label{ang23}
\end{align} 
where $V_{ij}$ are the elements of mixing matrix (\ref{V44}).
\section{Model Description}

\begin{table}
\centering
\begin{tabular}{|m{1.3cm}|m{0.2cm}m{0.2cm}m{0.2cm}m{0.3cm}|m{0.2cm}m{0.4cm}m{0.3cm}|m{0.4cm}|m{0.3cm}|m{0.3cm}|m{0.4cm}|m{0.4cm}|m{0.2cm}m{0.4cm}m{0.5cm}|m{0.4cm}|}
\hline
$\frac{Fields}{Charges}$ & $l$ & $e_R$& $\mu_R$& $\tau_R$ & H & H$^{\prime}$ &H$^{\prime\prime}$ & $\phi$ &$\psi$ &$\chi$& $\chi^{\prime}$ & $\zeta$ & $\nu_{R1}$&$ \ \nu_{R2} $&$\ \nu_{R3}$& S \\
\hline
$SU(2)_L$& 2 &1 &1 &1 & 2&2 & 2 &1 &1 &1 &1 &1 &1 &1 &1 &1  \\
$A_4$& 3 &1 &1$^{\prime\prime}$ &1$^{\prime}$ &1$^{\prime}$ &1 &1 &3 &3 &1 &1$^{\prime}$ &1 &1$^{\prime}$ &1 &1$^{\prime}$ &1$^{\prime\prime}$ \\
$Z_4$&1 &1 &1 &1 &1 &-i & i &-i &1 &1 &1 &-1 &i &-1 &1 &-i \\
$Z_2$&1 &1&1 &1 &1 &-1 &1 &1 &1 &1 &1 &1 &1 & -1&1 &1\\
\hline
\end{tabular}
\caption{Particle contents of the model and their group charges.}
\label{Table2}
\end{table}
This model uses the $A_4$ discrete group to develop the neutrino mass matrices along with $Z_4$ and $Z_2$ in normal hierarchy (NH) only. $A_4$ has four irreducible representations in which three are singlets $(1,1^{\prime},1^{\prime\prime})$ and one is triplet$(3)$\cite{Ishimori}. We take the SM charged lepton doublet $l$ as triplet under $A_4$ and the right-handed charged lepton singlets $(e_R, \mu_R, \tau_R)$ as singlets $(1, 1^{\prime\prime}, 1^{\prime})$ respectively. It is convenient to extend the SM Higgs $H$ by adding two more BSM Higgs  $H^{\prime}$ and $ H^{\prime\prime}$ which are singlets under $A_4$ to produce a mass model consistent with the current experimental data. We use two $A_4$ triplet flavons $\phi$ and $\psi$ for generating $M_L\ \mbox{and}\ M_D$ and two singlets $(\chi, \chi^{\prime})$ which will give the Majorana mass matrix. Another singlet flavon $\zeta$ is responsible for generating the sterile mass matrix $M_S.$ The full particle contents and their group charges are shown in Table $\ref{Table2}$. The invariant Lagrangian for the leptonic interactions is given by 
\begin{equation}
\mathcal{L} = \mathcal{L}_{M_L} + \mathcal{L}_{M_D} + \mathcal{L}_{M_R} + \mathcal{L}_{M_S} + h.c.
\end{equation}
where, 
\begin{equation}
\mathcal{L}_{M_L} = \frac{y_e}{\Lambda}(\overline{l}H\psi)_1e_R +\frac{y_{\mu}}{\Lambda}(\overline{l}H\psi)_{1^{\prime}}\mu_R + \frac{y_{\tau}}{\Lambda}(\overline{l}H\psi)_{1^{\prime\prime}}\tau_R
\end{equation}
\begin{equation}
\mathcal{L}_{M_D} =\frac{y_1}{\Lambda}(\overline{l}\tilde{H}\phi)_{1^{\prime\prime}}\nu_{R1} +\frac{y_2}{\Lambda}(\overline{l}\tilde{H^{\prime}}\phi)_{1}\nu_{R2} + \frac{y_3}{\Lambda}(\overline{l}\tilde{H}\psi)_{1^{\prime\prime}}\nu_{R3}
\end{equation}
\begin{equation}
\mathcal{L}_{M_R} = \frac{1}{2}\lambda_1\chi^{\prime}\overline{\nu}_{R1}^c\nu_{R1} + \frac{1}{2}\lambda_2\chi\overline{\nu}_{R2}^c\nu_{R2} +\frac{1}{2}\lambda_3\chi^{\prime}\overline{\nu}_{R3}^c\nu_{R3}
\end{equation}
\begin{equation}
\mathcal{L}_{M_S} = \frac{1}{2}k\zeta\overline{S}^c\nu_{R1}
\end{equation}
The constant $\Lambda$ denotes the cut-off scale and $\tilde{H} = i \tau_2 H $ ( where $\tau_2$ is the second Pauli matrix ) is used in order to make the Lagrangian guage invariant whereas $y_i,\ y_j\ ,\lambda_j,\ k$ where $i=e,\mu,\tau;\ j=1,2,3\ $ respectively, are the Yukawa coupling constants. $Z_2$ is used to remove the term $\frac{y_1}{\Lambda}(\overline{l}\tilde{H^\prime}\psi)_{1^{\prime\prime}}\nu_{R1}$ from $\mathcal{L}_{M_D}$. If we choose the $T-$diagonal basis of $A_4$ along with the flavon vev alignments\cite{Das1}
\begin{align}
\langle\phi\rangle =(v,0,0) \ \ \ \ ;\ \  \langle\psi\rangle=(v,v,v) \ \ \ ;\ \ \langle\chi\rangle = \langle\chi^{\prime}\rangle = v \ \  ;\ \ \langle\zeta\rangle= u.
\end{align} 
then, we get a diagonal charged lepton mass matrix 
\begin{equation}
M_L =\frac{\langle H\rangle v}{\Lambda} diag(y_e,\ y_{\mu},\ y_{\tau})
\end{equation}

The Dirac, Majorana and sterile mass matrices takes the following forms  

\begin{align}
M_D^o &= \frac{\langle H\rangle v}{\Lambda}\left(\begin{matrix}
y_1 & y_2 & 0 \\ 
y_1 & y_2 & 0 \\ 
y_1 & y_2 & y_3
\end{matrix}\right)\ \  =\ \ \left(\begin{matrix}
a & b & 0 \\ 
a & b & 0 \\ 
a & b & c
\end{matrix}\right);
\end{align}
\begin{align}
M_R &= v\left(\begin{matrix}
\lambda_1 & 0 & 0 \\ 
0 & \lambda_2 & 0 \\ 
0 & 0 & \lambda_3
\end{matrix}\right)\ \ =\ \ \left(\begin{matrix}
d & 0 & 0 \\ 
0 & e & 0 \\ 
0 & 0 & f
\end{matrix}\right);
\end{align}
\begin{align}
M_S &= \left(\begin{matrix}
k u & 0 & 0
\end{matrix}\right),
\end{align}
where $ a = \frac{\langle H\rangle v}{\Lambda}y_1$,$\ \ b = \frac{\langle H\rangle v}{\Lambda}y_2,\ \ c = \frac{\langle H\rangle v}{\Lambda}y_3,\ \ d =\lambda_1 v,\ \ e =\lambda_2 v, \ \ f =\lambda_3 v.$

Applying MES mechanism with these mass matrices in eq. $(\ref{mv})$, we obtain the active neutrino mass matrix as 
\begin{equation}
m_{\nu}^o = -\left(
\begin{array}{ccc}
 \frac{b^2}{e} & \frac{b^2}{e} & \frac{b^2}{e} \\
 \frac{b^2}{e} & \frac{b^2}{e} & \frac{b^2}{e} \\
 \frac{b^2}{e} & \frac{b^2}{e} & \frac{b^2}{e}+\frac{c^2}{f} \\
\end{array}
\right).
\end{equation}

It is easy to see that $m_{\nu}^o$ is a $\mu-\tau$ symmetric matrix which give $\theta_{13}=0$. But, recent experimental data has proven $\theta_{13}$ to be non-zero. In order to generate $\theta_{13}\neq 0$, $M_D^o$ is modified by adding a perturbation term $M_D'$. We can achieve this if we introduce an $SU(2)_L$ singlet flavon $\eta$ having $ A_4 \otimes Z_4 \otimes Z_2 $ charges $(3,-i,1)$ with vev alignment of $(0,0,v)$ in our model. The Lagrangian responsible for the perturbation matrix is 
\begin{equation}
\mathcal{L}_{M_D^{\prime}} = \frac{y_4}{\Lambda}(\bar{l}\eta)_{1'}\tilde{H}\nu_{R1} +\frac{y_4}{\Lambda}(\bar{l}\eta)_{1}\tilde{H'}\nu_{R2} + \frac{y_4}{\Lambda}(\bar{l}\eta)_{1^{\prime\prime}}\tilde{H^{\prime\prime}}\nu_{R3}
\end{equation}
Then, the perturbation matrix looks like
\begin{equation}
M_D'= \frac{\langle H\rangle v}{\Lambda}\left(\begin{matrix}
0 & 0 & y_4 \\ 
0 & y_4 & 0 \\ 
y_4 & 0 & 0
\end{matrix}\right).
\end{equation}

 The resultant active neutrino mass matrix with $M_D=M_D^o + M_D'\ $ from eq. $(\ref{mv})$ becomes 
 \begin{equation}
 m_{\nu} \simeq -\left(
\begin{array}{ccc}
 \frac{b^2}{e}+\frac{t^2}{f} & \frac{b (b+t)}{e} & \frac{b^2}{e}+\frac{c t}{f} \\
 \frac{b (b+t)}{e} & \frac{(b+t)^2}{e} & \frac{b (b+t)}{e} \\
 \frac{b^2}{e}+\frac{c t}{f} & \frac{b (b+t)}{e} & \frac{b^2}{e}+\frac{c^2}{f} \\
\end{array}
\right),
\end{equation} 
 where $t=\frac{\langle H\rangle v}{\Lambda}y_4,$ and $y_4$ is the Yukawa coupling for the perturbation term.

The sterile neutrino mass is obtained from eq. $(\ref{ms})\ $ as 
\begin{equation}
m_s \simeq -\left(
\begin{array}{c}
 \frac{g^2}{d} \\
\end{array}
\right),
\end{equation}
 where $g=k u$.

The numerical bounds of the model parameters and mixing parameters obtained from our model will be determined in the next section through numerical analysis.

\section{Numerical Analysis and Results}
\begin{figure}[H]
\begin{minipage}{0.45\textwidth}
\includegraphics[scale=0.7]{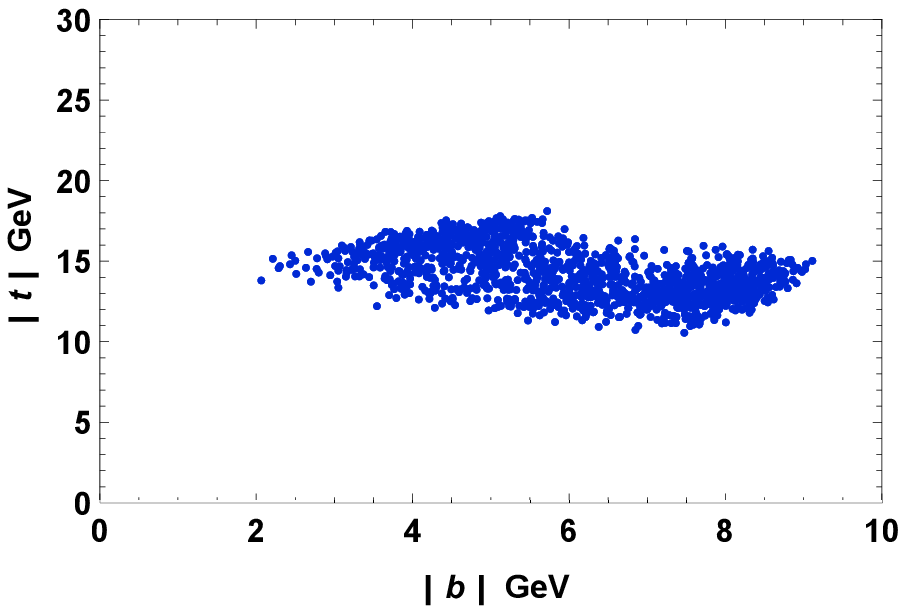} 
\end{minipage}
\hspace{1cm}
\begin{minipage}{0.45\textwidth}
\includegraphics[scale=0.7]{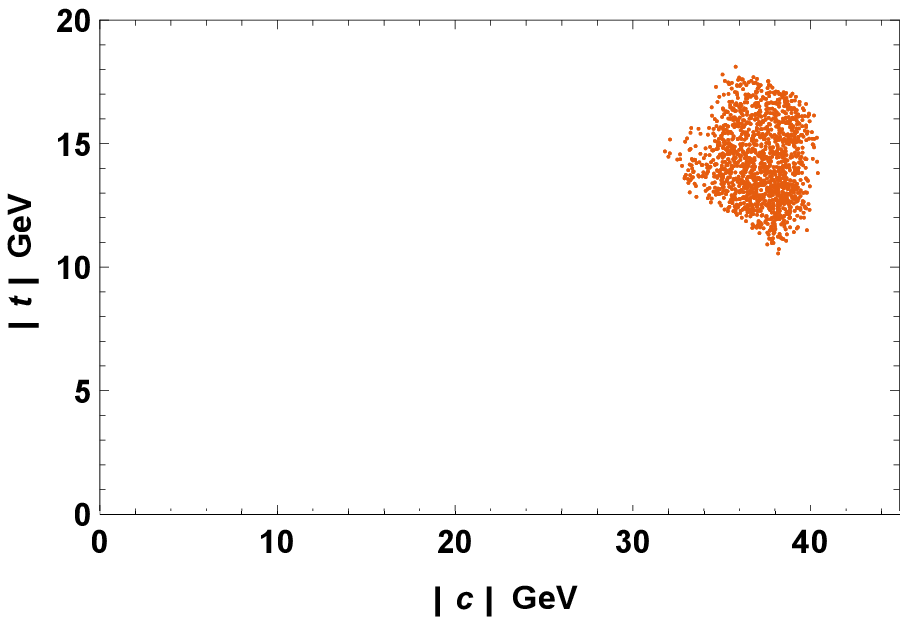} 
\end{minipage}
\caption{\footnotesize Variation of model parameters are shown as correlation plots with each other.}
\label{parameter}
\end{figure}

To validate the present model, we first try to solve the free parameters by comparing the LHS and RHS of eq. $(\ref{u33})$. We take the current 3$\sigma$ values of mixing angles, and mass squared differences from Table $\ref{Table1}$. We vary the unknown Majorana phases in the range $(0,2\pi)$ and we have fixed non-degenerate values for the heavy right-handed neutrino mass parameters $d=e\simeq 10^{13}$ GeV and $f\simeq 5\times 10^{13}$ GeV. We numerically solve the model parameters $b,c$ and $t$ which satisfy the five independent equations with a tolerance of $\mathcal{O}(10^{-2})$. The correlation plots among different models parameters are shown in Fig. $\ref{parameter}$. We find that the parameter space is very narrow, which can be verified or discarded in future experiments.

\begin{figure}[H]
\begin{minipage}{0.45\textwidth}
\includegraphics[scale=0.7]{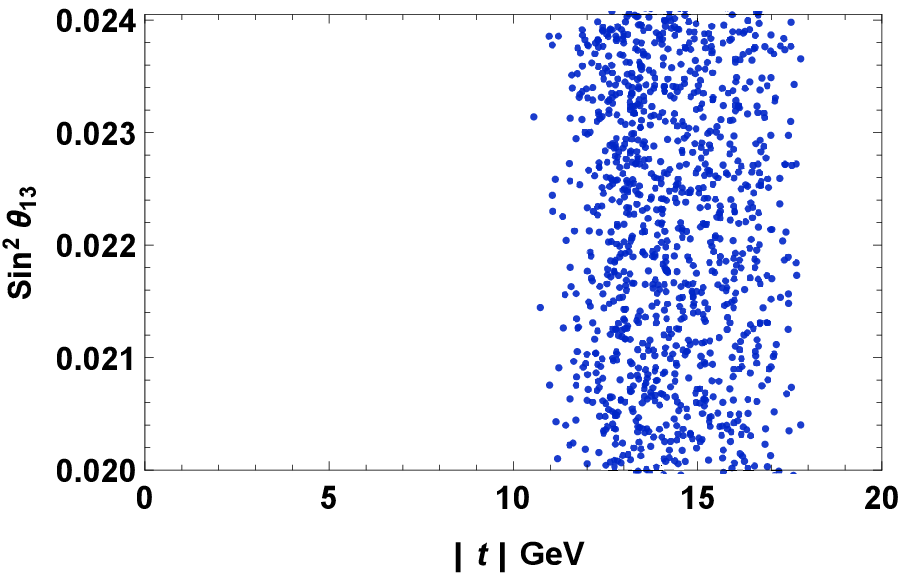} 
\end{minipage}
\hspace{1cm}
\begin{minipage}{0.45\textwidth}
 \includegraphics[scale=0.7]{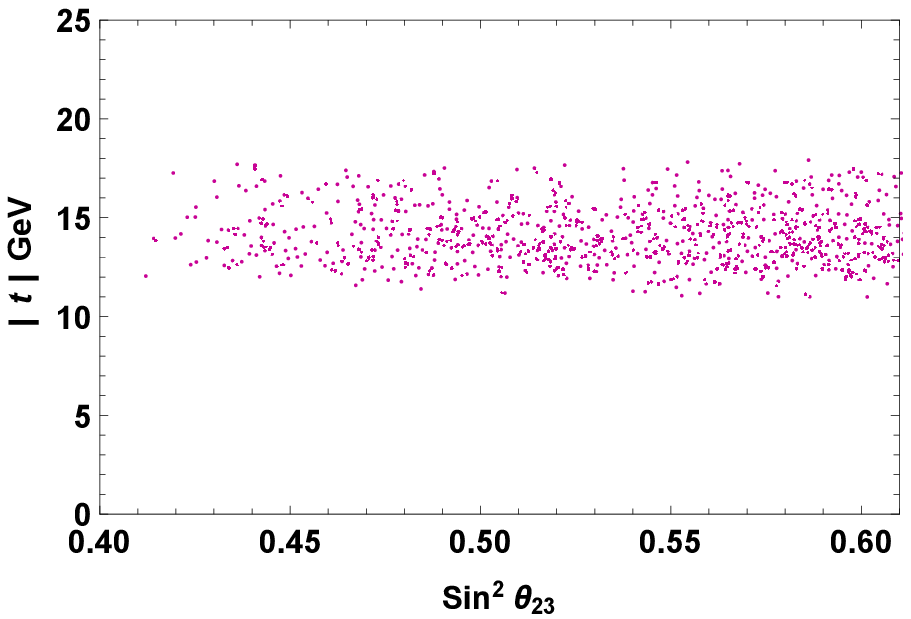} 
\end{minipage}
\label{mixingangle}
\caption{\footnotesize Variation of perturbation parameters $t$ with mixing angles.}
\label{tvsangle}
\end{figure}

\begin{figure}[H]
\begin{minipage}{0.45\textwidth}
 \includegraphics[scale=0.7]{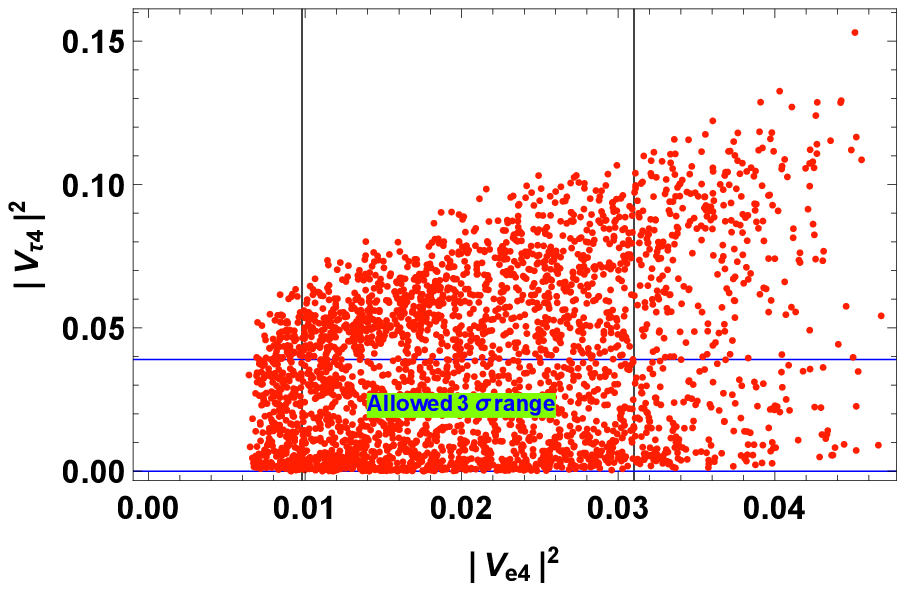} 
\end{minipage}
\hspace{1cm}
\begin{minipage}{0.45\textwidth}
\includegraphics[scale=0.7]{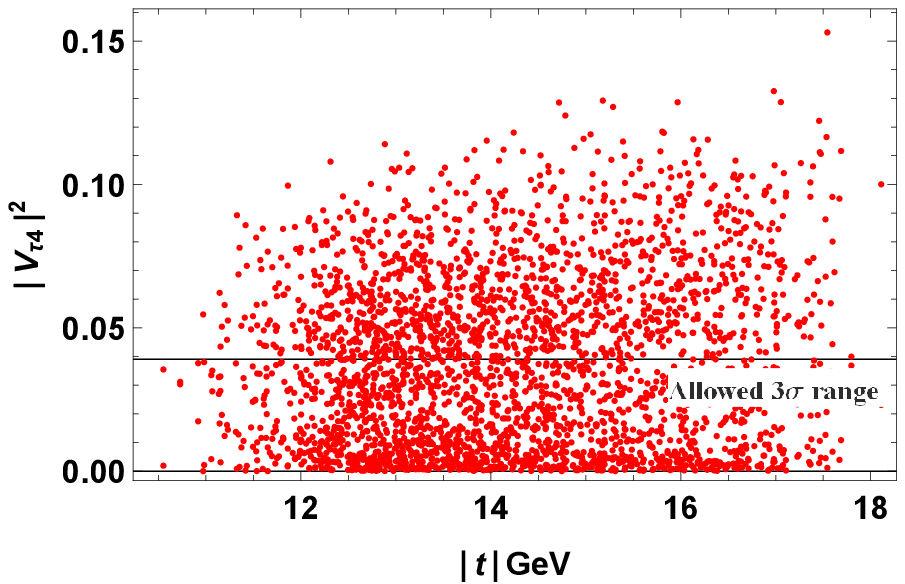} 
\end{minipage}
\caption{\footnotesize Variation of active-sterile mixing element $|V_{e4}|^2$ with $|V_{\tau 4}|^2$ and with model parameter $t$. }
\label{actste}
\end{figure}

Remaining parameters $a$ and $g$ are solved using the 3$\sigma$ bounds on active-sterile mass-squared difference $|\Delta m_{41}^2|\in (0.87,2.04)~$ eV$^2$\cite{Katrin2020,goswami,2021planck}. Now, the model-dependent $4\times 4$ mixing matrix $V$ can be developed in eq. $(\ref{V44})$ from which we can solve other mixing parameters using eq. $(\ref{ang14})$ - eq. $(\ref{ang23})$. The variation of different mixing angles with perturbation parameter $t$ are shown in Fig. \ref{tvsangle}. It is observed that many data points of $t$ are available within the 3$\sigma$ ranges of the mixing angles. More data points are concentrated at regions $\sin^2\theta_{23}>0.50$ which implies that the model favours higher octant of $\theta_{23}$.  Fig. \ref{actste} shows the variation of active-sterile mixing elements. Allowed 3$\sigma$ bounds are shown in the plots and it can be seen that the present model can give values within the experimental bounds.

\begin{table}[H]
\begin{center}
\begin{tabular}{|c|c|c|}
\hline 
Parameters & 3$\sigma$ range (GeV) & Best-fit (GeV) \\ 
\hline 
$|b|$ & 2.05 - 9.11 & 4.45 \\ 
$|c|$ & 32.11 - 40.35 & 35.70 \\ 
$|t|$ & 10.75 - 18.41 & 17.01 \\ 
$|a|$ & 9.56 - 21.03 & 14.68 \\ 
$|g|$ & 96.66 - 119.53 & 106.77\\
\hline
\end{tabular} 
\caption{3$\sigma$ range and the best-fit values of the model parameters.}
\label{Table3}
\end{center}
\end{table}

\newpage

In case where mixing variables in eq. $(\ref{u33})$ take best-fit values from Table $\ref{Table1}$ and taking Majorana phases to be zero for simplicity, we determine the model parameters and they are shown in Table $\ref{Table3}$. The resulting best-fit $4\times 4$ mixing matrix is obtained as 
\begin{equation*}
\small
U_{bf}=\left(
\begin{array}{cccc}
 0.7969 & 0.5584 & 0.1665 & 0.1375 \\
 -0.2980+0.0211 i & 0.6011\, +0.0154 i & -0.7075-0.1811 i & 0.1375 \\
 0.6788\, +0.0719 i & -0.5711+0.0129 i & -0.5858-0.1497 i & 0.2968 \\
 -0.2879-0.0282 i & 0.0107\, -0.0040 i & 0.2644\, +0.0755 i & 0.9370 \\
\end{array}
\right)
\end{equation*}

We can observe that the elements of the mixing matrix are well within the allowed ranges of oscillation data.

\begin{figure}[H]
\begin{minipage}{0.45\textwidth}
\includegraphics[scale=0.7]{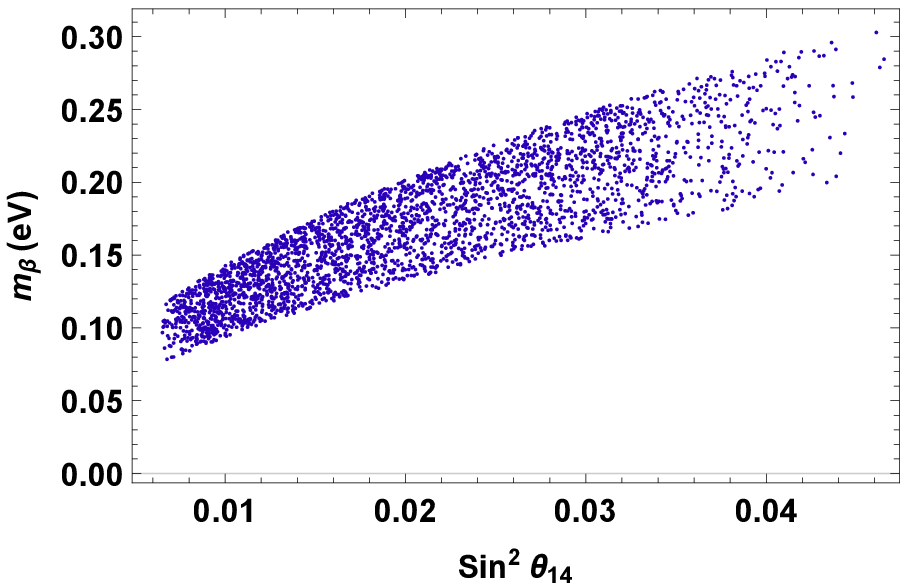}  
\end{minipage}
\hspace{1cm}
\begin{minipage}{0.45\textwidth}
\includegraphics[scale=0.7]{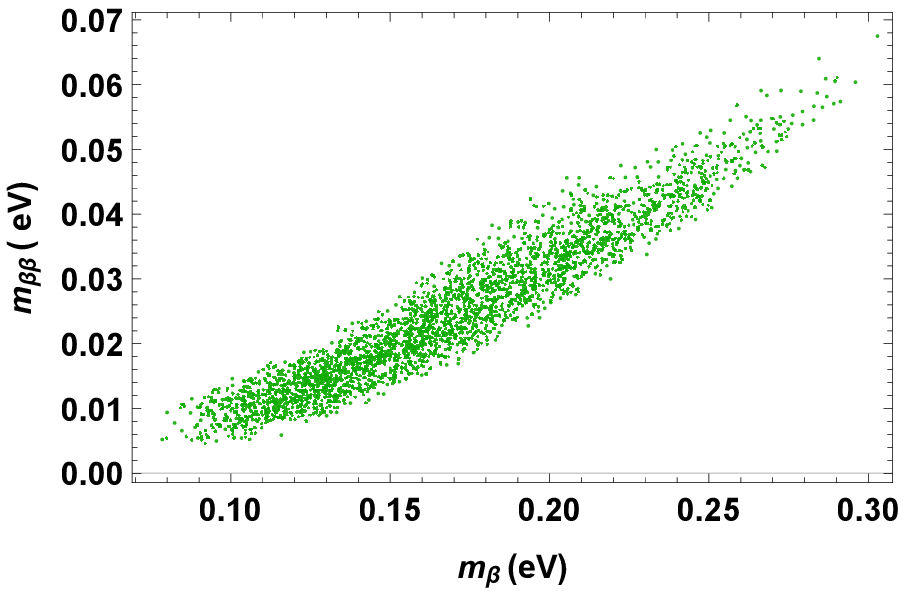}   
\end{minipage}
\caption{\footnotesize Variation of effective neutrino mass $m_{\beta}$ and $ m_{\beta\beta}$ with active-sterile mixing angles}
\label{meff}
\end{figure}

We also calculate the effective neutrino mass $m_{\beta}$  and  $m_{\beta\beta}$ and plot their variations in Fig. \ref{meff}. We find that the mixing angles and effective neutrino mass parameters are within their allowed ranges.

\begin{figure}[H]
\begin{minipage}{0.45\textwidth}
\includegraphics[scale=0.7]{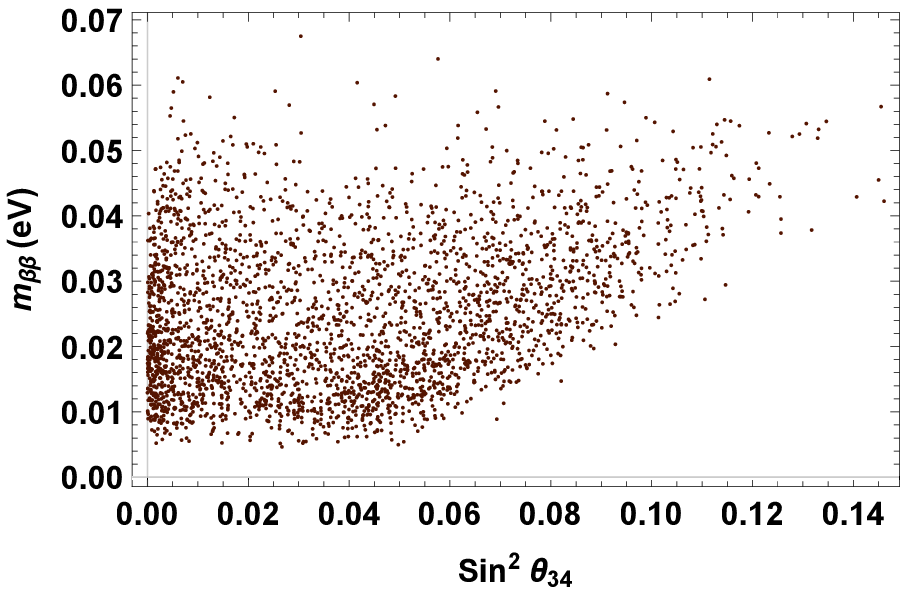}  
\end{minipage}
\hspace{1cm}
\begin{minipage}{0.45\textwidth}
\includegraphics[scale=0.7]{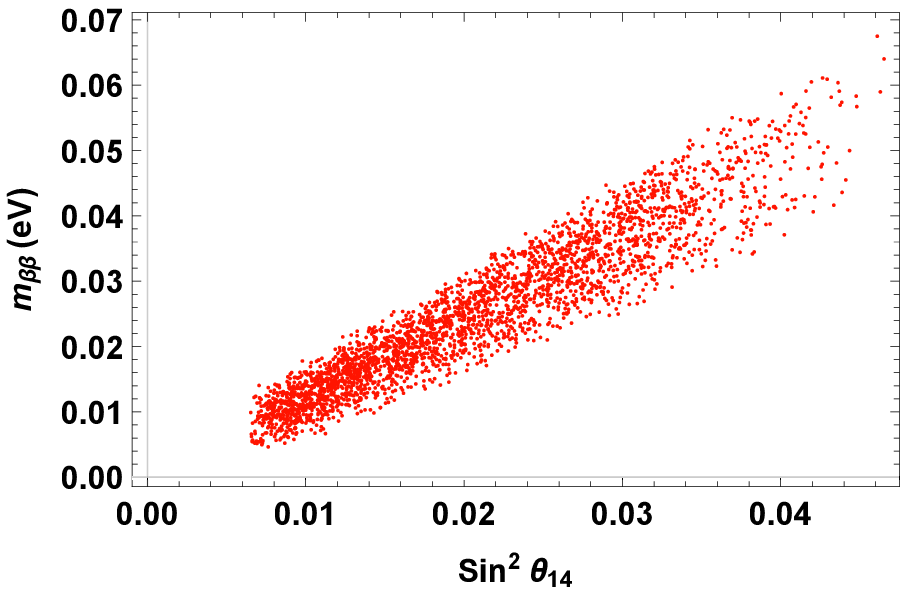}  
\end{minipage}
\end{figure}
\section{Summary and discussion}
In this work, we have developed an $A_4$ model supplemented by $Z_4$ and $Z_2$ groups. One singlet sterile neutrino is added to the 3-neutrino theory to explain $3+1$ active sterile neutrino masses and mixings. Addition of triplet flavon $\eta$ gives the desired $\mu-\tau$ symmetry breaking active neutrino mass matrix. The active sterile mixing matrix R provides the non-unitary contribution to the active mixing matrix $U_{PMNS}.$ By constraining the light neutrino masses $m_1,m_2,m_3$ and $m_4$ from the experimental mass-squared difference at 3$\sigma$, we have determined the parameters of the model. It is observed that the parameters lie in the low GeV scale, which can be verified in future experiments. We have plotted the bounds on the active sterile mixing $|V_{\mu 4}|^2$ and $|V_{\tau 4}|^2$ from our model by fixing $|\Delta m_{41}^2|$ and $|V_{e4}|^2$. A large number of data points  are concentrated within the allowed ranges. We also calculate the effective mass parameters $m_{\beta}$ and $m_{\beta\beta}$. Their values are obtained in the ranges $0.0784 ~ \mbox{eV} < m_{\beta} < 0.3034$ eV and $0.00342~\mbox{eV} < m_{\beta\beta} < 0.06675$ eV respectively. This is allowed in the latest upper bound $m_{\beta}<1.1$ eV at 90$\%$ confidence level recently published by the KATRIN Collaborations\cite{aker2019improved}. This is also in agreement with the results of the analysis in Ref. \cite{Hagstot}, where active neutrinos mixing with light sterile sterile neutrino leads to an upper limit of $m_{\beta}<0.09$ eV and  $m_{\beta\beta}<0.07$ eV at 95$\%$ CL.  

At current status, the bounds on active-sterile mass squared difference $|\Delta m_{41}|^2$ is still not known. Many experiments give various constraints, such as, $|\Delta m_{41}|^2 =1.7$eV$^2$\cite{gariazzo2017updated}, $|\Delta m_{41}|^2 =4.5$eV$^2$\cite{icecube2020}, $|\Delta m_{41}|^2 < 10$eV$^2$\cite{dayabay2020}, $|\Delta m_{41}|^2 =7.3 \pm 1.17$eV$^2$\cite{serebrov2021analysis}, etc. We have chosen a particular bound and performed the numerical analysis. We have also determined the best-fit $4\times 4$ mixing matrix resulting from the model and found that the mixing parameters satisfy the current 3$\sigma$ bounds of the mixing matrix. In conclusion, we can remark that we have developed a model framework which is possible to explain the origin of neutrino masses and mixings through discrete symmetry. It can also generate possible active-sterile mixing in the 3+1 MES mechanism.

\section*{Acknowledgements}
We wish to thank Prof. M. K. Das of Tezpur University for giving us many fruitful suggestions in this work. One of us (MKS), would also like to thank DST-INSPIRE, Govt. of India, for providing financial support under DST-INSPIRE Fellowship. 

\bibliographystyle{unsrt}
\bibliography{activesterilemixing}
\end{document}